\documentclass[12pt]{iopart}
\usepackage{graphicx}
\usepackage{amssymb}
\begin{document}

\title[KPZ dynamics from a variational perspective]{KPZ dynamics from a variational perspective:\\potential landscape, time behavior, and other issues}

\author{H S Wio$^1$, M A Rodr\'{\i}guez$^1$, R Gallego$^2$, R R Deza$^3$ and J A Revelli$^4$}
\address{$^1$ IFCA (Cantabria U. \& CSIC), Avda.\ de los Castros s/n, E-39005 Santander, Spain}
\address{$^2$ Math.\ Dept., Oviedo U. (Gij\'on Campus), E-33203 Gij\'on, Spain}
\address{$^3$ IFIMAR (Mar del Plata U. \& CONICET), De\'an Funes 3350, B7602AYL Mar del Plata, Argentina}
\address{$^4$ FaMAF-IFEG (C\'ordoba U. \& CONICET), Av.\ Medina Allende s/n, X5016LAE C\'ordoba, Argentina}
\ead{wio@ifca.unican.es, rodrigma@ifca.unican.es, rgallego@uniovi.es, rdeza@ifimar-conicet.gob.ar, revelli@famaf.unc.edu.ar}
\date{\today}

\begin{abstract}
The deterministic KPZ equation has been recently formulated as a gradient flow, in a \emph{nonequilibrium potential} (NEP)
\[\Phi[h(\mathbf{x},t)]=\int\mathrm{d}\mathbf{x}\left[\frac{\nu}{2}(\nabla h)^2-\frac{\lambda}{2}\int_{h_0(\mathbf{x},0)}^{h(\mathbf{x},t)}\mathrm{d}\psi(\nabla\psi)^2\right].\]
This NEP---which provides at time \(t\) the \emph{landscape} where the stochastic dynamics of \(h(\mathbf{x},t)\) takes place---is however \emph{unbounded}, and its exact evaluation involves \emph{all} the detailed histories leading to \(h(\mathbf{x},t)\) from some initial configuration \(h_0(\mathbf{x},0)\). After pinpointing some consequences of these facts, we study the time behavior of the NEP's first few moments and analyze its signatures when an external driving force \(F\) is included. We finally show that the asymptotic form of the NEP's time derivative \(\dot\Phi[h]\) turns out to be valid for \emph{any} substrate dimensionality $d$, thus providing a valuable tool for studies in $d>1$.
\end{abstract}

\vspace{2pc}

Submitted to: \emph{J. Stat. Mech.}
\maketitle

\section{Introduction}

The KPZ equation for kinetic interface roughening (KIR) \cite{kpz,HHZ,BarSta,Krug},
\begin{equation}\label{eq:1}
\partial_th(\mathbf{x},t)=\nu\,\nabla^2h(\mathbf{x},t)+\frac{\lambda}2\left[\nabla h(\mathbf{x},t)\right]^2+\xi(\mathbf{x},t),
\end{equation}
where \(h(\mathbf{x},t)\) is the interface height and \(\xi(\mathbf{x},t)\) a Gaussian noise with
\[\langle\xi(\mathbf{x},t)\rangle=0\:,\:\langle\xi(\mathbf{x},t)\xi(\mathbf{x}',t')\rangle=D\,\delta(\mathbf{x}-\mathbf{x}')\,\delta(t-t'),\] is nowadays a paradigm of systems exhibiting nonequilibrium critical scaling \cite{necs}. In fact---besides standing out as a representative of a large and robust class of microscopic KIR models, from which the phenomenological parameters $\nu$, $\lambda$, $D$ can be computed---it is intimately related to two highly nontrivial problems:
\begin{enumerate}
\item Through the Hopf--Cole transformation \[\phi(x,t)=\exp\left[\frac{\lambda}{2\nu}h(x,t)\right],\]
Eq.\ (\ref{eq:1}) is isomorphic to the diffusion equation with multiplicative noise obeyed by the restricted partition function of directed polymers in random media (DPRM). Thus the DPRM problem belongs to the KPZ universality class, and its progress reinforces that of KIR, as much as \emph{vice versa}.
\item Via \(\mathbf{v}=-\nabla h\), Eq.\ (\ref{eq:1}) can be mapped (for $\lambda=1$) into the Burgers equation for a randomly stirred vorticity-free fluid \cite{burgers1,burgers2}. As a consequence (being the nonlinear term in the latter, part of the substantial derivative), $\lambda$ has to be \emph{invariant} under scale changes. The invariance of $\lambda$ under scale changes leads to the remarkable relation
\begin{equation}\label{eq:2}
\alpha+z=2
\end{equation} 
(a signature of the KPZ universality class) in \emph{any} substrate dimensionality $d$.
\end{enumerate}

Since the first and third terms in the l.h.s.\ of Eq.\ (\ref{eq:1}) were already present in the model by Edwards and Wilkinson (EW), the innovation came from the tilt-dependent local growth velocity in the second term (or rather, by its interplay with the third one). Under additive uncorrelated local Gaussian white noises, plane (eventually moving) interfaces---which are stable for \(\xi(\mathbf{x},t)=0\)---develop (in both EW and KPZ models and for large enough systems) into statistically self-affine fractals, whose typical roughness width scales as a power \(\beta\) (called the \emph{growth} exponent) of the elapsed time. \(\beta\) is in turn the ratio between the interface's H\"older (or \emph{roughness}) exponent \(\alpha\) and the \emph{dynamic} exponent \(z\), governing the growth in time of the correlation length. In the well understood $d=1$ case, \(\alpha\) turns out to be that of simple random walk (namely \(\alpha=\frac1{2}\)) in both models. However, Eq.\ (\ref{eq:2}) imposes that \(z\) (and thus \(\beta\)) be appreciably different (at least after some crossover time, needed for the second term to take over the first). The initial success of Eq.\ (\ref{eq:1}) was then due to the consistency of \(\beta_\mathrm{KPZ}\) with KIR experimental results.

Equation (\ref{eq:2}) is often attributed to the Burgers equation's Galilean invariance, which in turn translates into KPZ equation's invariance under changes in tilt. This opinion has been repeatedly challenged and in fact, the numerically computed exponents obey Eq.\ (\ref{eq:2}) in a discrete version of Eq.\ (\ref{eq:1}) where both Galilean invariance and the (1d--peculiar) fluctuation--dissipation theorem are explicitly broken \cite{Nos-1,Nos-2}.

By about half its lifetime so far, the field was mature enough to approach not only the second moment of the fluctuations in \(h\) but its full statistics. The innovation began within the field of DPRM, and led to propose for \(h(x,t)\) the asymptotic behavior \cite{prsp0a,prsp0b} (see also \cite{SaSp,CaLeD,takeuchi})
\begin{equation}\label{eq:3}
h(x,t)\sim v_\infty\,t+(\Gamma t)^{1/3}\,\chi(x'),
\end{equation}
with \[A:=\frac\nu{2D}\:,\:\Gamma:=A^2\frac{\lambda}2\mbox{ , and
 }x':=\frac{Ax}2(\Gamma t)^{-2/3}.\]
The stochastic variable \(\chi\) obeys the Tracy--Widom (TW) statistics of the largest eigenvalue of a random-matrix ensemble, here determined by the geometry of the substrate \footnote{The Gaussian orthogonal ensemble (GOE) if  the substrate is flat, the Gaussian unitary (GUE) one if it is curved.}. In the stationary state however, temporal correlations are governed by the Baik--Rains (BR) $F_0$ limit distribution \cite{baik-rains}, not related to random-matrix theory (RMT).

In the last few years, a handful of exact solutions to Eq.\ (\ref{eq:1}) have arisen: whereas most of them were inspired by RMT \cite{SaSp,CaLeD,rmt1,rmt2}, one was born right within the field of stochastic differential equations \cite{hairer}. On the other hand, the field is by now so mature that experiments can decide on the statistics \cite{takeuchi,tasa10,taea,tasa12}. And pretty much the same occurs with the numerics: in a recent review article, the statistics of KPZ itself has been found to agree with the results transposed from DPRM \cite{hahe12,hahe13,hata15}.

It is the purpose of this article to lighten up the abovementioned developments from the perspective of a recent variational formulation of the KPZ equation \cite{wio-01,Nos-1,Nos-2,Nos-3,pip}. In the following section we derive general consequences from some properties of the nonequilibrium potential functional $\Phi[h]$. Next, we undertake a numerical study of the time dependence of $\Phi[h]$ and derive some consequences from the forced case. Then we relate the time behavior of $\dot\Phi[h]$ with the asymptotic form in Eq.\ (\ref{eq:3}) and finally, we collect our conclusions.

\section{The functional}

Some deeply rooted folklore (but nothing else than that) denied that the KPZ system---exhibiting as it does such a complex behavior---could ever be expressed as a stochastic gradient flow, namely
\begin{equation}\label{eq:4}
\partial_th(\mathbf{x},t)=-\frac{\delta\Phi[h]}{\delta h(\mathbf{x},t)}+\xi(\mathbf{x},t).
\end{equation}
As shown in Ref.\ \cite{wio-01}, the trouble is not that Eq.\ (\ref{eq:4}) is not valid but that the functional \(\Phi[h]\)---defined by
\begin{equation}\label{eq:5}
\Phi[h(\mathbf{x},t)]=\int\mathrm{d}\mathbf{x}\left[\frac{\nu}{2}(\nabla h)^2-\frac{\lambda}{2}
\int_{h_0(\mathbf{x},0)}^{h(\mathbf{x},t)}\mathrm{d}\psi(\nabla\psi)^2\right]
\end{equation}
with \(h_0(\mathbf{x},0)\) an \emph{arbitrary} initial pattern, usually assumed to be constant (in particular, \(h_0=0\))---has \emph{not} an explicit density. Even though at time \(t\), \(\Phi\) depends \emph{only} on the field \(h(\mathbf{x},t)\), its evaluation requires \emph{knowing the detailed history} that led from \(h_0\) to \(h(\mathbf{x},t)\). In other words, retrieving information on \(\Phi\) (such as its landscape at certain time or the time dependence of its mean value) requires averaging not simply over field configurations \(h(\mathbf{x})\) at time \(t\), but over (statistically weighted) \emph{trajectories} of the field configuration.

Being the KPZ equation a stochastic gradient flow, its NEP \(\Phi\) (which governs its \emph{deterministic} component) provides the \emph{landscape} where the stochastic dynamics of \(h(\mathbf{x},t)\) takes place, and fulfils explicitly the Lyapunov property \(\dot{\Phi}[h]=-\left[\frac{\delta\Phi[h]}{\delta h(x,t)}\right]^2\leq0\) \footnote{Having the KPZ equation no stationary state, \(\Phi\) is to be regarded as a \emph{generalization} of the classical definition of a ``nonequilibrium potential'' \cite{graham}.}. Unfortunately, this does \emph{not} make \(\Phi\) into a Lyapunov functional, since it is unbounded from both above and below. In fact, a formal Taylor expansion \(\hat\Phi[h]\) of \(\Phi[h]\) around some reference pattern \(h_0\) yields a \emph{cubic} effective potential \(\hat\Phi(h)\), since the $n$--th variation \(\delta^n\Phi\equiv0, n>3\). This fact lends itself to insightful interpretations.

\subsection{On normalization}

A known analogy---the equilibrium particle density distribution in a constant gravitational field---will help clarify this point. In principle, it would be the stationary limit of the (exact) solution of the Fokker--Planck equation (FPE) associated to the Langevin problem \[\dot x=F+\xi(t),\quad\langle\xi(t)\rangle=0,\quad\langle\xi(t)\xi(t')\rangle=D\,\delta(t-t'),\] with the initial condition \(P(x,t_0|x_0,t_0)=\delta(x-x_0)\), if this stationary distribution could be normalized! As known, the physics of this simple problem dictates that there is always a boundary. In fact, the correct interval has a \emph{finite} limit in the sense of the decreasing potential, thus allowing for normalization.

The FPE associated to Eq.\ (\ref{eq:1}) is \cite{kpz,HHZ}
\begin{eqnarray}
\partial_tP[h]&=&\int\mathrm{d}x\,\frac{\delta}{\delta h}\left[-\left(\nu\nabla^2h+\frac{\lambda}{2}(\nabla h)^2\right)P+D\frac{\delta P}{\delta h}\right]\nonumber\\
&=&\int\mathrm{d}x\,\frac{\delta}{\delta h}\left[\left(\frac{\delta\Phi}{\delta h}\right)P+D\frac{\delta P}{\delta h}\right].\label{eq:6}
\end{eqnarray}
In this case, it is not a \emph{stationary} but an \emph{asymptotic} solution to Eq.\ (\ref{eq:6}) what we look for. Forcing the condition \(\partial_tP[h]=0\) leads to \(P_\mathrm{as}[h]\propto\exp(-\Phi[h]/D)\). As given by Eq.\ (\ref{eq:5}), \(\Phi[h]\) is a multidimensional potential function. Nonetheless a projection of it---the effective potential \(\hat\Phi(h)\), obtained through the expansion of the functional in Eq.\ (\ref{eq:12})---will exhibit a simple cubic-like shape. As in the simple analogy before, this pdf cannot be normalized except by considering an interval \((-\infty,h_m]\).

\subsection{On the statistics}

Assuming the referred cubic-like shape for the NEP and at least in 1d (where according to recent exact results the pdf is not a Gaussian, but a Tracy--Widom distribution) we may reason as follows:
\begin{itemize}
\item at the left of the small barrier, the NEP increases with negative \(h\) values, preventing the occurrence of large \(h<0\), and so the left branch of the distribution should decay at least like a Gaussian;
\item on the other hand, even before the small barrier has been overcome, large \(h>0\) are slightly more probable than \(h<0\), and thus the right branch should decay much slowly.
\end{itemize}
Together with the results of \cite{prsp0a,prsp0b,SaSp,CaLeD}, this observation indicates that the well known pdf \cite{HHZ,BarSta}
\begin{equation}\label{eq:7}
P_\mathrm{stat}[h]\propto\exp\left\{-\frac\nu{2D}\int(\nabla h)^2\mathrm{d}x\right\},
\end{equation}
makes sense only for a finite 1d system with periodic boundary conditions and for
times larger than \(t \sim L^z\), the saturation one!

\subsection{Non-Markov character of the KPZ dynamics}

We can write the space discrete version of Eq.\ (\ref{eq:5}) as
\begin{equation}\label{eq:8}
\Phi[h]=\sum_{j=1}^L\Delta x\left[\frac{\nu}{2}(\partial_x h_j)^2-\frac{\lambda}{2}
\int_{h_j^0}^{h_j}\mathrm{d}\psi(\partial_x\psi_j)^2\right]
\end{equation}
where the index \(j=1,2,\ldots,L\) indicates the lattice site. The lattice spacing is usually adopted as \(\Delta x=1\). In order to evaluate the second term we resort to the following approximation,
\begin{eqnarray}
\int_{h_j^0}^{h_j}\mathrm{d}\psi(\partial_x\psi)_j^2\:&&\approx\sum_{\mu=0}^{M-1}\tau\left(\frac{h_{j,\mu+1}-h_{j,\mu}}\tau\right)(\partial_x\tilde{h}_{j,\mu})^2\nonumber\\
&&\approx\int_0^t\mathrm{d}s\,\dot{h}(x,s)[\partial_x h(x,s)]^2,\label{eq:9}
\end{eqnarray}
with \(\tau\) the time step, \(\mu\) the time index, \(\tilde{h}_{j,\mu}\) some intermediate value between \(h_{j,\mu}\) and \(h_{j,\mu+1}\), and  \(\dot{h}(x,s):=\lim_{\tau\to0}(h_{j,\mu+1}-h_{j,\mu})/\tau\). This allows us to write the potential in the form
\begin{equation}\label{eq:10}
\Phi[h]=\int\mathrm{d}x\left\{\frac{\nu}{2}(\nabla h)^2-\frac{\lambda}{2}
\int_0^t\mathrm{d}s\,\dot{h}(x,s)[\partial_x h(x,s)]^2\right\}
\end{equation}
which highlights the ``non Markov'' character of the KPZ dynamics. Moreover, since the potential value depends on the whole trajectory, it really implies a very long (``infinite'') memory. It seems natural to immediately relate this fact with the ageing phenomena found in relation with KPZ \cite{agng1,agng2}.

\subsection{Expansion}

To close this section we recall that the potential \(\Phi[h]\) can be expanded in a
Taylor-like form \cite{pip} starting from a reference pattern \(h_0\), using \(h(x)=h_0(x)+\delta h(x)\), according to
\begin{equation}\label{eq:11}
\hat\Phi[h]=\Phi[h_0]+\delta\Phi[h_0]+\frac1{2}\delta^2\Phi[h_0]+\frac1{6}\delta^3\Phi[h_0]+\ldots
\end{equation}
However, for all \(n\geq4\) we find \(\delta^n\Phi[h_0]\equiv0\) (\(\forall h_0\)). Hence, the above indicated form reduces to
\begin{equation}\label{eq:12}
\hat\Phi[h]=\Phi[h_0]+\delta\Phi[h_0]+\frac1{2}\delta^2\Phi[h_0]+\frac1{6}\delta^3\Phi[h_0].
\end{equation}
If we consider a \(h_0\) such that \(\delta\Phi[h_0]=0\) (for instance \(h_0=0\)), call \(u(x,t)=\delta h(x)\) and separate the reference potential values according to \(\hat\Phi[h]=\Phi[h_0]+\hat\Phi[h]\), we find \cite{pip}
\begin{equation}\label{eq:13}
\hat\Phi[h]=\int\mathrm{d}x\left(\frac{\nu}{2}-\frac{\lambda u}{6}\right)(\partial_x u)^2,
\end{equation}
a form that clearly shows the existence of a \emph{diffusive instability}, and justifies our previous argument on the NEP having a cubic-like form. However, this expression is \emph{only an approximation}, as it seems to correspond to evaluating \(\Phi[h]\) along a \emph{single} (albeit highly probable) trajectory.

The above indicated expansion allows us to evaluate the relative ``stability'' of
two different patterns. The ``potential energy'' difference between patterns \(h_1\) and \(h_2=h_1+\varepsilon\) is
\begin{eqnarray}
\Delta\Phi[h]&&=\Phi[h_2]-\Phi[h_1]\nonumber\\
&&=\hat\Phi[\varepsilon]=\int\mathrm{d}x\left(\frac{\nu}{2}-\frac{\lambda\varepsilon}{6}\right)(\partial_x\varepsilon)^2\label{eq:14}\\
&&\approx\sum\Delta x\left(\frac{\nu}{2}-\frac{\lambda\varepsilon_j}{6}\right)\frac1{2}\left[(\varepsilon_{j+1}-\varepsilon_j)^2(\varepsilon_j-\varepsilon_{j-1})^2\right]\nonumber
\end{eqnarray}
where we have used the discrete form for \((\partial_x\varepsilon)^2\) exploited in \cite{Nos-1,Nos-2}. This makes the NEP's cubic polynomial character even more apparent.

\begin{figure}[htbp]
\begin{center}
\includegraphics[width=.7\columnwidth,bb= 0pt 0pt 354.6pt 700.2pt]{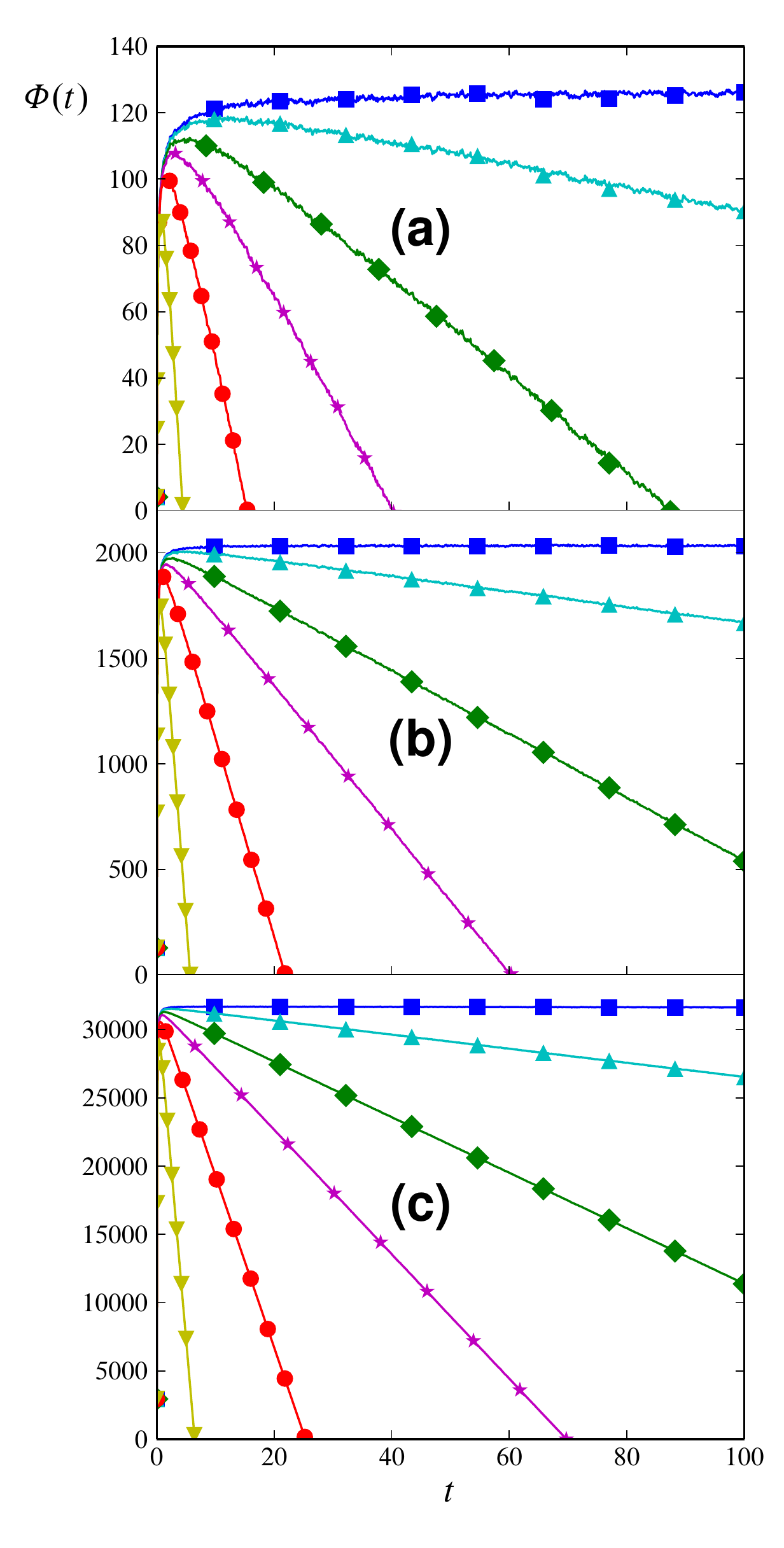}
\caption{Time behavior of \(\Phi[h]\), averaged over 100 samples, in (a) 1d (size 1,024), (b) 2d (size \(128^2\)), and (c) 3d (size \(64^3\)). \(\blacksquare:\,\lambda=0.01\), \(\blacktriangle:\,\lambda=0.10\), \(\diamond:\,\lambda=0.20\), \(\bigstar:\,\lambda=0.30\), \(\bullet:\,\lambda=0.50\), \(\blacktriangledown:\,\lambda=1.00\). Since the statistics past this maximum is not Gaussian, particular histories dominate the mean.}
\label{fig:1}
\end{center}
\end{figure}

\begin{figure}[htbp]
\begin{center}
\includegraphics[width=.7\columnwidth,bb= 0pt 0pt 354.6pt 700.2pt]{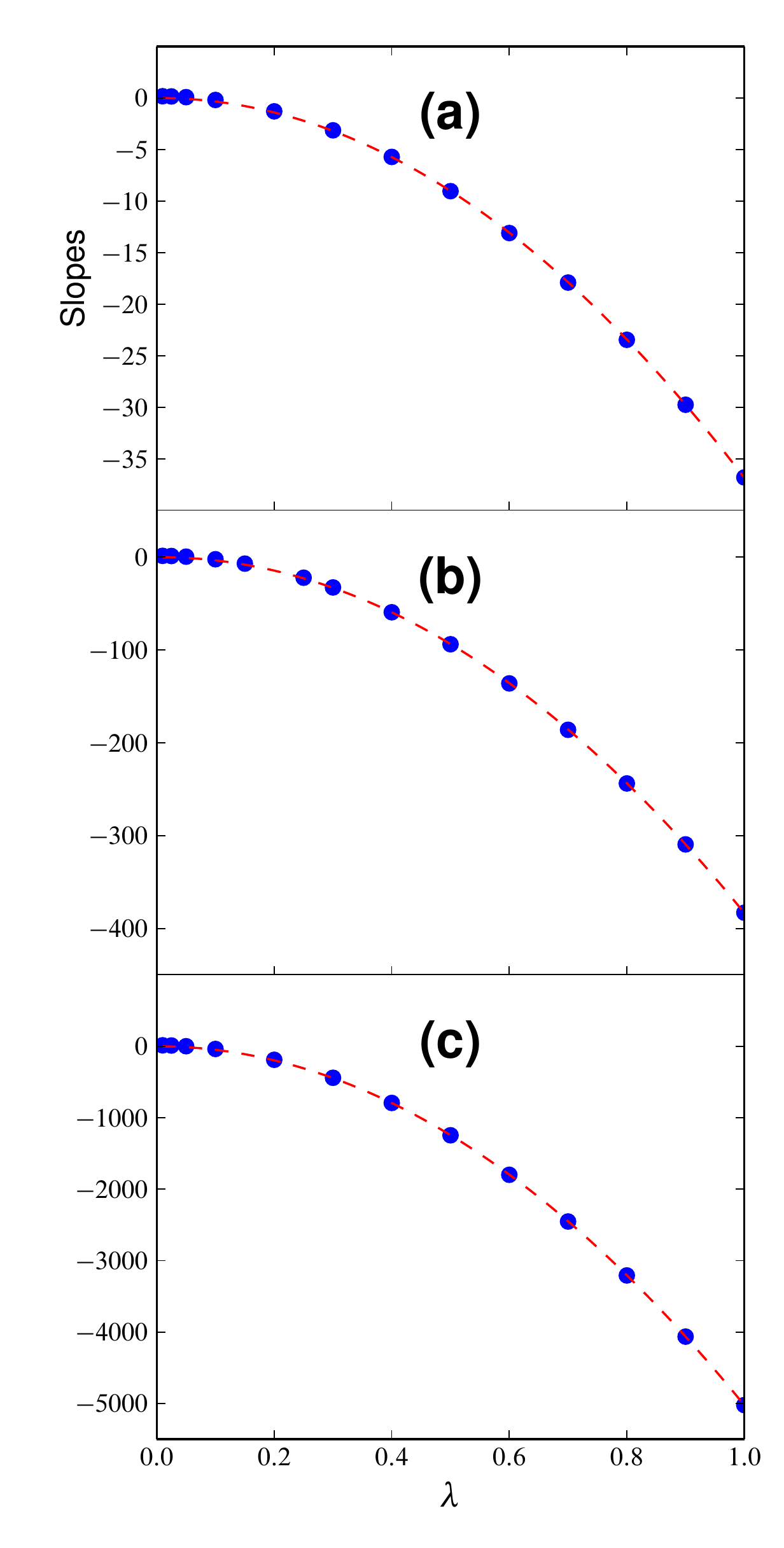}
\caption{Mean \(\lambda\)-dependence over 100 samples of the NEP's asymptotic slope, in (a) 1d (size 1,024), (b) 2d (size \(128^2\)), and (c) 3d (size \(64^3\)). Dashed lines: best fits with \(a\,\lambda^b\) yield \(b=2.01\) in 1d and 2d, and \(b=2.03\) in 3d.}
\label{fig:2}
\end{center}
\end{figure}

\begin{figure}[htbp]
\begin{center}
\includegraphics[width=.8\columnwidth,bb= 0pt 0pt 821pt 635pt]{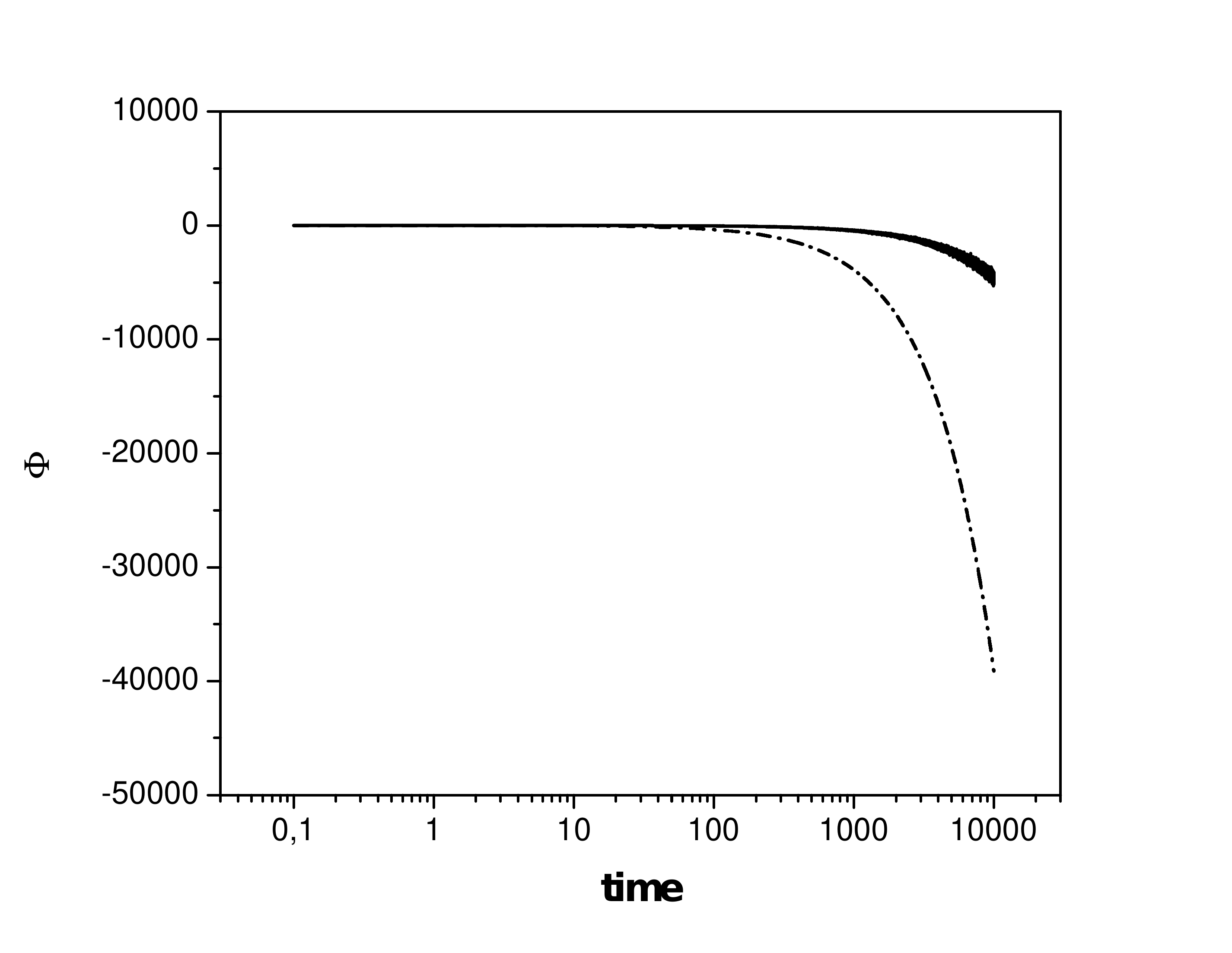}
\caption{Comparison of \(\Phi(t)\) from Eq.\ (\ref{eq:5})---solid line---with \(\hat\Phi(t)\) from Eq.\ (\ref{eq:12})---dashed line---for \(\lambda=1\) in 1d. Both graphs collapse when one of them is multiplied by a proper scale factor.}
\label{fig:3}
\end{center}
\end{figure}

\section{Numerical results for \(\Phi[h]\)}

Using a spectral method to integrate Eq.\ (\ref{eq:1}), we have analyzed the time behavior of the first few moments of \(\Phi[h]\): its \emph{mean}, \emph{dispersion}, \emph{skewness} and \emph{kurtosis}. Partially due to a better treatment of the $\nabla h$ term, spectral methods have proved to be more stable and reliable than finite-differences schemes in the integration of some nonlinear growth equations \cite{gacl07,gigr02,gall11}. Figure \ref{fig:1} displays the time dependence of the NEP's average over 100 samples, for systems in 1d (size 1,024), 2d (size \(128^2\)), and 3d (size \(64^3\)), and several values of \(\lambda\). For any \(\lambda>0\) there is a maximum, where the nonlinear (KPZ) term overcomes the linear (EW) one. Past this maximum, \(\langle\Phi[h(t)]\rangle_t\sim A-Bt\) (with \(B\sim\lambda^2\) as explicit in Fig.\ \ref{fig:2}). This result shows that---due to the correlations and in an effective way---the NEP behaves as having only a \emph{linear} dependence on \(h\) (just as in the toy model discussed before).

Figure \ref{fig:3} shows that the \(\log t\) behavior of \(\hat\Phi(t):=\langle\hat\Phi[h(t)]\rangle_t\) is qualitatively similar to that of \(\Phi(t):=\langle\Phi[h(t)]\rangle_t\) (there is only a relative shift), and both graphs collapse when one of them is multiplied by a proper scale factor.

\begin{figure}[htbp]
\begin{center}
\includegraphics[width=.7\columnwidth,bb= 0pt 0pt 441pt 369pt]{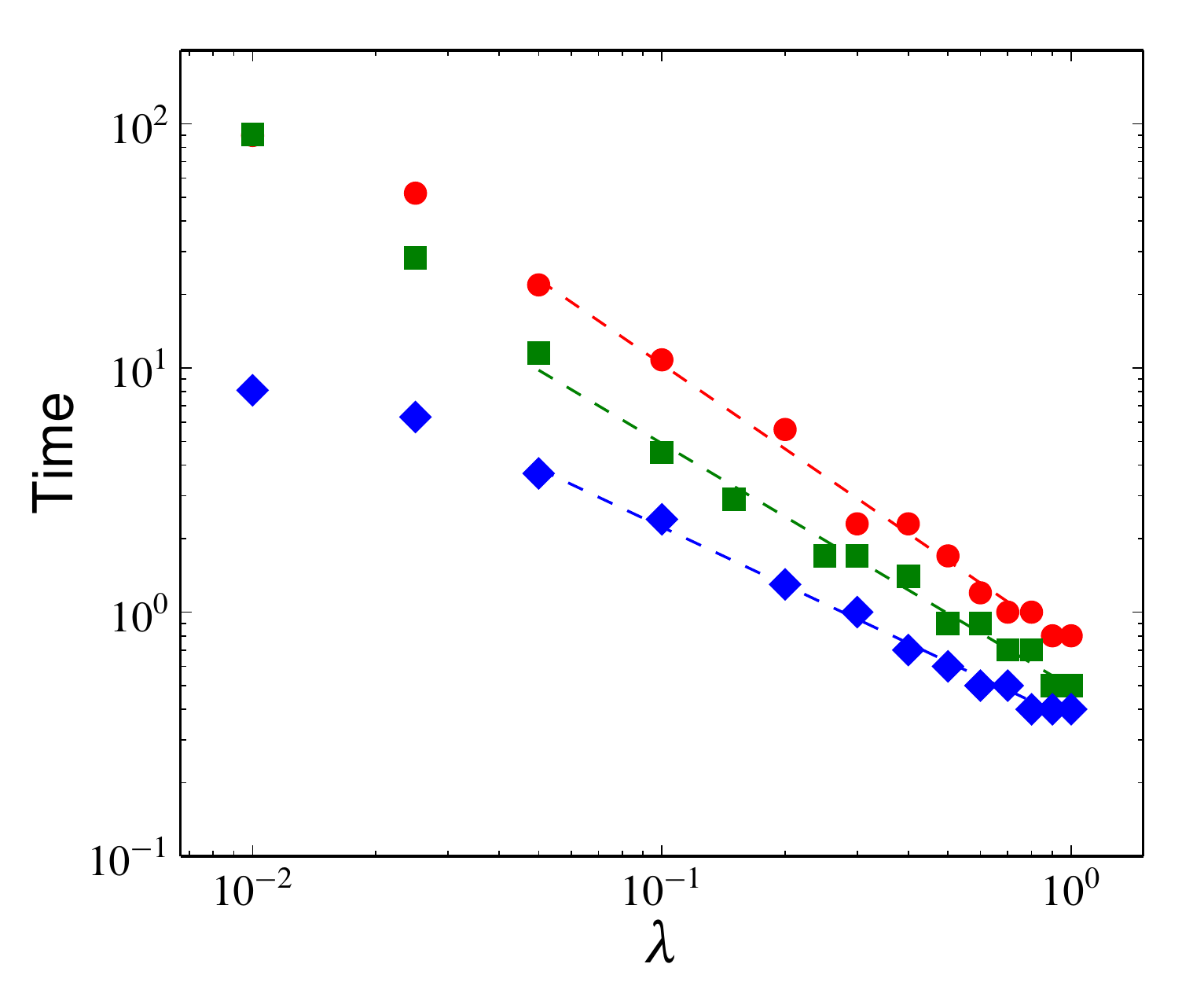}
\caption{Mean \(\lambda\)-dependence over 100 samples of the time of occurrence of the NEP's maximum, in \(\bullet\): 1d (size 1,024), \(\blacksquare\): 2d (size \(128^2\)), and \(\diamond\): 3d (size \(64^3\)). Dashed lines: best fits with \(a\,\lambda^{-b}\) yield \(b=1.14\) in 1d, \(b=0.98\) in 2d, and \(b=0.79\) in 3d.}
\label{fig:4}
\end{center}
\end{figure}

\begin{figure}[htbp]
\begin{center}
\includegraphics[width=.7\columnwidth,bb= 0pt 0pt 354.6pt 700.2pt]{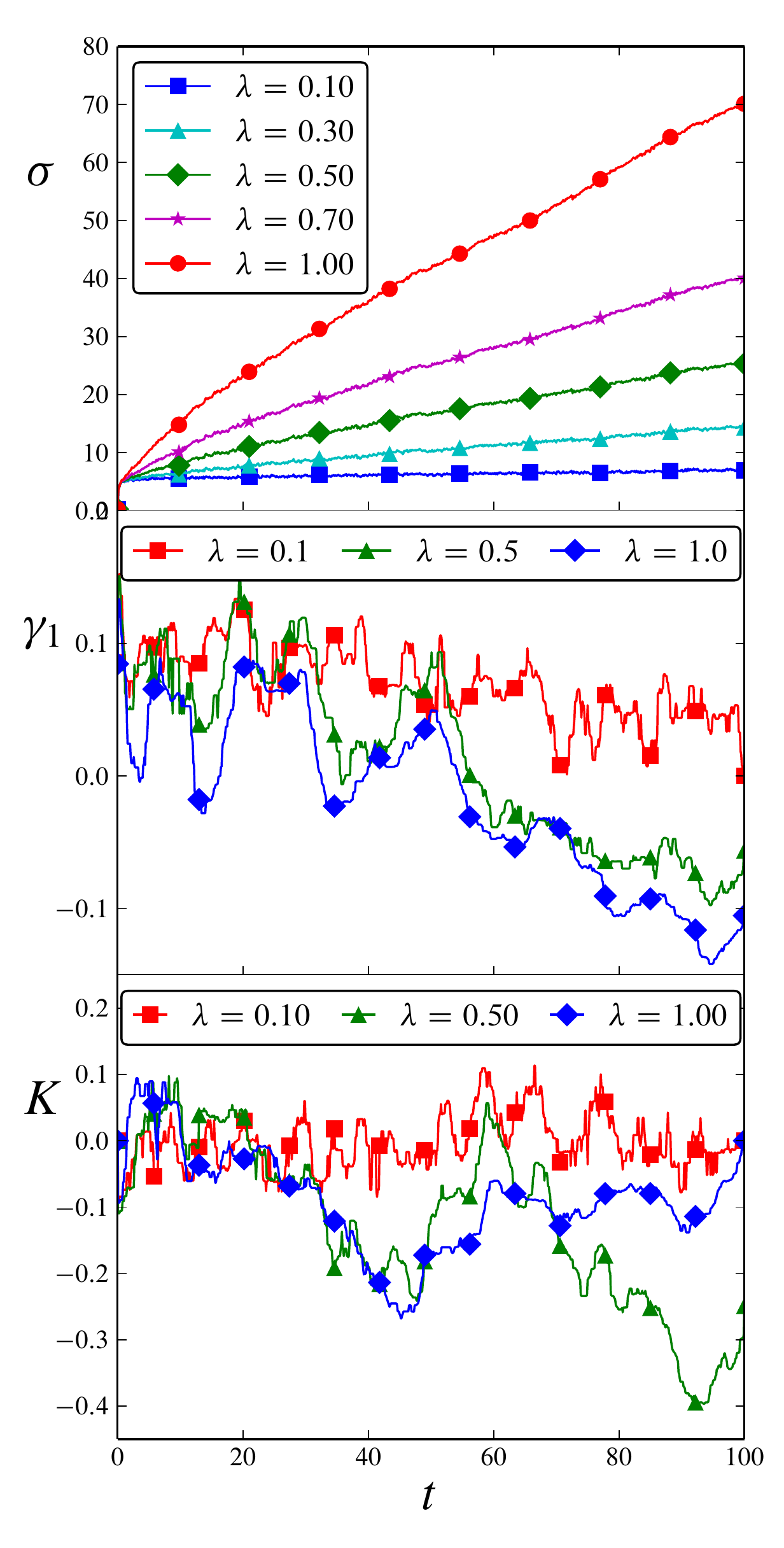}
\caption{Time behavior of the NEP's \emph{dispersion} \(\sigma\), \emph{skewness} \(\gamma_1\) and \emph{kurtosis} \(K\) in 1d, from 1,000 samples of size 1,024.}
\label{fig:5}
\end{center}
\end{figure}

The dependence on \(\lambda\) of the time at which the maximum occurs---a proxy of the EW-to-KPZ crossover time \cite{FoTo,g3,olea}---is shown in Fig.\ \ref{fig:4}. Although roughly compatible with a \(t_\mathrm{xovr}\sim\lambda^{-1}\) law (in agreement with the results of previous studies exploiting the time behavior of the stochastic action \cite{appb,pi-kpz}), this dependence becomes milder as d increases, supporting the conjecture of the existence of an upper critical dimension.

Figure \ref{fig:5} displays for the 1d case, the time behavior of the NEP's \emph{dispersion} \(\sigma(t):=\langle\{\Phi[h(t)]-\Phi(t)\}^2\rangle_t^{1/2}\), \emph{skewness} \(\gamma_1(t):=\langle\{\Phi[h(t)]-\Phi(t)\}^3\rangle_t/\sigma^3\) and \emph{kurtosis} \(K(t):=\langle\{\Phi[h(t)]-\Phi(t)\}^4\rangle_t/\sigma^4-3\) \footnote{We have performed a median filter on the kurtosis and skewness in order
to smooth out the curves, which aids to analyze the data. By doing this,
the important information is captured, leaving out fine-scale phenomena.}. In the observed time interval, \(\sigma\) appears to increase continuously, without hints of saturation; a detailed analysis of the data shows an initial dependence of the form \(t^{3/4}\), followed by a crossover to a dependence \(t^{1/2}\). At least for \(\lambda\) strong enough, \(\gamma_1\) seems to drift from positive to negative values; this can be understood by considering the cubic-like shape of the potential: (i) for short times, the pdf is concentrated in the metastable left well---with a very small probability of large \(h<0\) values---while there is a long tail of \(h>0\) values; (ii) for long times, the pdf essentially concentrates outside the shallow metastable well, with only a short tail for \(h<0\) values. Finally, the (much noisier) behavior of \(K\) seems to indicate that the distribution of \(\Phi[h]\) has a \emph{peak}, that looks similar to the Gaussian case.

\begin{figure}[htbp]
\begin{center}
\includegraphics[width=.65\columnwidth,bb= 0pt 0pt 441pt 657pt]{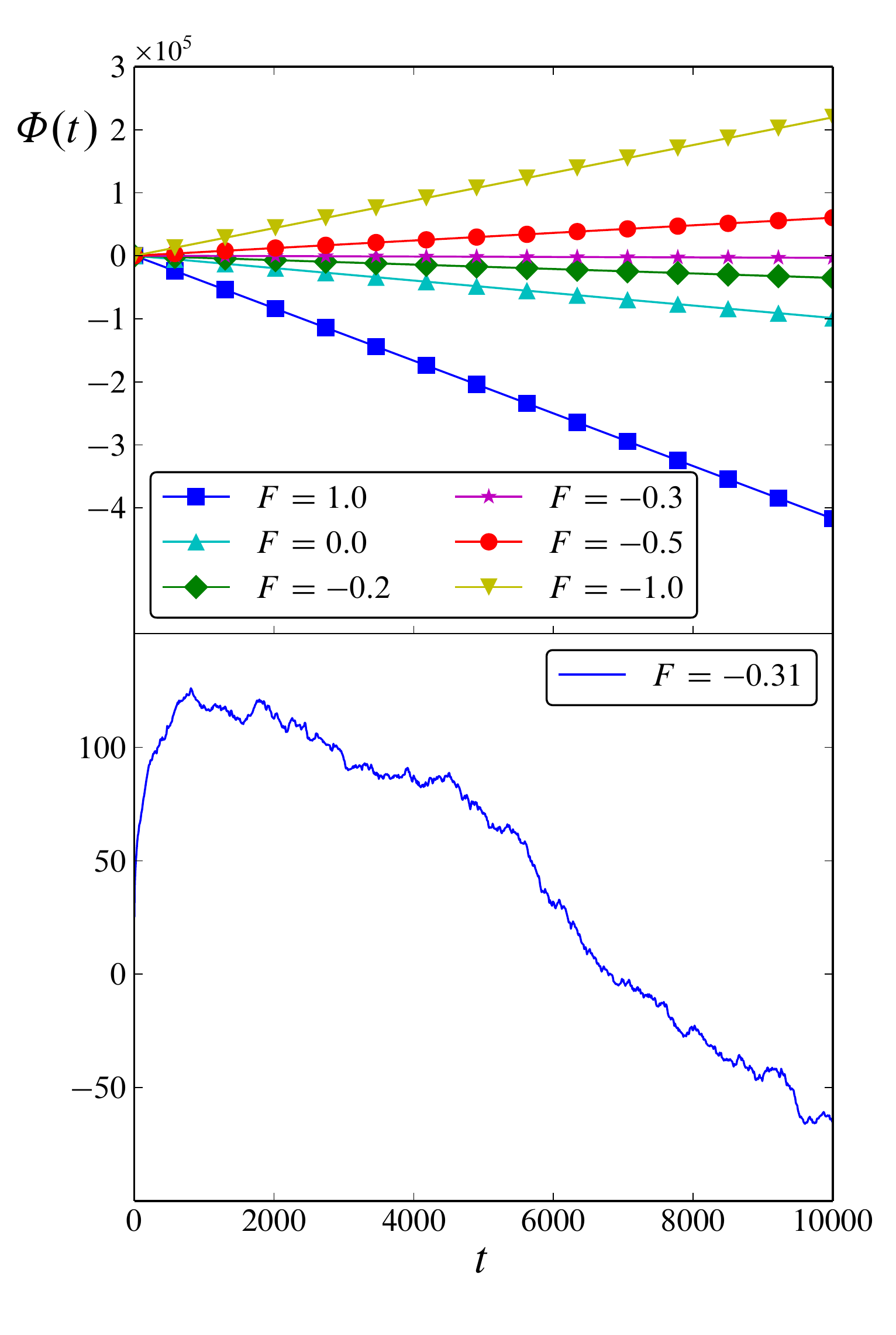}
\caption{NEP's long-time behavior in 1d (\(\lambda=1\)), from 100 samples of size 256. (a) For \(F<-0.3\), \(\langle\Phi[h]\rangle\) \emph{increases} with time in the observation interval; (b) detail showing that for \(F\) \emph{just slightly} lower than -0.3, \(\langle\Phi[h]\rangle\) first \emph{increases} with time (on a much longer timescale than the initial rise due to the EW term) and then begins to \emph{decrease}, supporting the hypothesis of an \emph{activation-like} behavior.}
\label{fig:6}
\end{center}
\end{figure}

\subsection{External field}

By including an external driving force \(F\), it is possible to capture some new aspects of the dynamics. Figure \ref{fig:6} shows (on a much longer timescale than in Fig.\ \ref{fig:1}, so that the already discussed initial rise due to the EW term cannot be appreciated) the NEP's time behavior for both positive and negative values of \(F\). The \(F>0\) case brings no novelty; but when negative enough forces are applied, a \emph{reversion} is observed of the NEP's time behavior, corresponding to a \emph{reversion of the front motion} \(\langle h\rangle\) and strongly supporting the hypothesis of an \emph{activation-like} behavior (which could be guessed from the cubic-like form of the NEP commented before). However, this activation-like behavior is \emph{not} of the exponential (Kramers-like) form and moreover, it is valid only for a limited range of \(F\). This assertion can be understood by looking at the form of the NEP,
\begin{eqnarray}
\Phi[h]&&=\int\mathrm{d}x\left[\frac{\nu}{2}(\nabla h)^2-\frac{\lambda}{2}
\int_0^h\mathrm{d}\psi(\nabla\psi)^2-F\,h(x,t)\right]\nonumber\\
&&\approx\int\mathrm{d}x\left[\left(\frac{\nu}{2}-\frac{\lambda u}{6}\right)(\partial_x u)^2-F\,u\right].\label{eq:15}
\end{eqnarray}
If \(F<0\), the evolution towards negative values of \(h\) proceeds mainly along the directions where \((\partial_x h)^2\approx(\partial_x u)^2=0\). However, as soon as---due to the presence of noise---\(h\) (or \(u\)) slightly depart from such (very low probability) directions, the effect of the NEP is such that it will drive the system in the opposite direction! However, this will only happen for \(F\) larger than some threshold value. This aspect will be fully clarified in the next Section. Thus for \(F=0\), the time behaviors of \(\langle\Phi[h]\rangle\) and \(\langle h\rangle\) are \emph{correlated}, even though the dispersion of the NEP (Fig.\ \ref{fig:5}) and that of the front differ (the former continuously increases with time, while the latter saturates for \(t\ge L^z\)).

\section{Some asymptotic results}

The average \(\langle\int\mathrm{d}x\,(\partial_x h)^2\rangle\) saturates to a constant \(C_0\) at long times. Using in Eq.\ (\ref{eq:15}) the discrete expression for \(\Phi[h]\)---Eq.\ (\ref{eq:8}), with the approximation in Eq.\ (\ref{eq:9})---we can write
\begin{eqnarray}
\langle\Phi[h]\rangle&&=\nu\,C_0-\frac{\lambda}2\left\langle\int\mathrm{d}x\int\mathrm{d}\psi\,[\partial_x\psi]^2\right\rangle-F\langle h\rangle\nonumber\\
&&\approx\nu\,C_0-\frac{\lambda}2\left\langle\sum_\mathrm{space}\sum_\mathrm{time}\delta h\,[\partial_x h]^2\right\rangle-F\langle h\rangle.\label{eq:16}
\end{eqnarray}
In the discrete representation, Eq.\ (\ref{eq:3}) implies \(\delta h\sim v_\infty\,\tau\) at long times, where \(\tau\) corresponds to \(\delta t\) and \(v_\infty\) is the asymptotic front velocity. Inserting this expression into the previous one and replacing \(\langle\int\mathrm{d}x\,(\partial_x h)^2\rangle\) by its bound \(C_0\), we get
\begin{equation}\label{eq:17}
\langle\Phi[h]\rangle=\nu\,C_0-(\lambda\,C_0+F)\,v_\infty\,t.
\end{equation}
This result completely agrees with the numerical ones presented above. In particular, \(\langle\Phi[h]\rangle\) will decrease at long times \emph{only if} \(\lambda\,C_0+F\ge0\). If \(\lambda\,C_0+F<0\), \(\langle\Phi[h]\rangle\) will \emph{increase} asymptotically, as shown in Fig.\ \ref{fig:6}.

In addition, it is well known that \(v_\infty\propto\lambda\) \cite{BarSta}. Hence (particularly in the \(F=0\) case) we find the \(\lambda^2\) dependence for the slope of \(\langle\Phi[h]\rangle\) \emph{vs.} \(t\), as obtained in simulations.

In order to close this section, let us look at the asymptotic behavior from another point of view, focusing in the \(F=0\) case. As already indicated, the statistics of the stochastic variable \(\chi\) in Eq.\ (\ref{eq:3}), is of paramount importance \cite{prsp0a,prsp0b}. Let us work in the continuous representation. In addition to the forms of \(\Phi[h]\)
given by Eqs.\ (\ref{eq:5}) and (\ref{eq:10}), we can also obtain an expression for \(\dot\Phi[h]\) as
\begin{equation}\label{eq:18}
\dot\Phi[h]=-\int\mathrm{d}x\left[\nu\nabla^2h+\frac{\lambda}2(\nabla h)^2\right]\dot{h}.
\end{equation}
We have the following relations
\begin{eqnarray}
\partial_xh&=&(\Gamma t)^{1/3}\partial_{x'}\chi\left(\frac{\mathrm{d}x'}{\mathrm{d}x}\right)\sim\frac{A}2(\Gamma t)^{1/3}(\Gamma t)^{-2/3}\chi'\nonumber\\
&\sim&\frac{A}2(\Gamma t)^{-1/3}\chi',\mbox{ with }\chi':=\partial_x\chi.\label{eq:19}
\end{eqnarray}
The time derivative of Eq.\ (\ref{eq:3}) is
\begin{equation}\label{eq:20}
\dot{h}\sim v_\infty+\frac{\Gamma}3(\Gamma t)^{-2/3}\,\chi-\frac{\Gamma}3Ax(\Gamma t)^{-4/3}\,\chi'\sim v_\infty,
\end{equation}
as the 2nd and 3rd terms decay very fast. On the other hand,
\[(\partial_xh)^2\sim\frac{A^2}4(\Gamma t)^{-2/3}\,(\chi')^2\:,\:\partial_x^2h\sim\frac{A^2}4(\Gamma t)^{-2/3}\,\chi''.\]
Exploiting these relations we obtain
\begin{equation}\label{eq:21}
\dot\Phi[h]\sim-\int\mathrm{d}x'\frac{A}2\left[\nu\chi''+\frac{\lambda}2(\chi')^2\right]v_\infty.
\end{equation}
If we look at long times, where the nonlinear contribution dominates, we have
\begin{equation}\label{eq:22}
\dot\Phi[h]\sim-\int\mathrm{d}x'\frac{A\lambda}4(\chi')^2v_\infty.
\end{equation}
Since \(v_\infty\propto\lambda\), we get right away the expected result: the constant goes as \(\lambda^2\).

\section{Conclusions}

The KPZ equation can be expressed as a gradient flow by means of the functional \(\Phi\) in Eq.\ (\ref{eq:5}). However, evaluating \(\Phi[h(\mathbf{x},t)]\) requires averaging not only over the ensemble of field configurations at time \(t\), but over their \emph{whole} respective stories, given that they started at \(h_0\). This implies that (even being a state functional) \(\Phi[h]\) has \emph{very long memory}, which gives us a clue on the occurrence of \emph{ageing} processes \cite{agng1,agng2}.

Even though it fulfils the Lyapunov property, the functional form of \(\Phi\) is that of a cubic polynomial and is thus unbound. The KPZ equation results to be a high-dimensional (non Kramers) escape problem, and its asymptotic pdf cannot be normalized on the whole \(h\)--space. We have argued moreover that the observed asymptotic Tracy--Widom statistics can be plausibly understood in terms of \(\Phi\)'s cubic shape.

Regarding \(\Phi\)'s time behavior, we have shown both numerically (Figs.\ \ref{fig:1} and \ref{fig:2}) and analytically in Eq.\ (\ref{eq:17}) that for \(F=0\), \(\langle\Phi[h]\rangle\sim A-Bt\), with \(B\sim\lambda^2\), implying that---due to effect of correlations---\(\Phi[h]\) acquires an effective linear dependence on \(h\).

As a proxy of the crossover time from the EW regime to the KPZ one, we have studied the dependence on \(\lambda\) of the time at which the maximum occurs, and found that it goes roughly as \(t_\mathrm{xovr}\sim\lambda^{-1}\), in agreement with previous results obtained by exploiting the time behavior of the stochastic action \cite{appb,pi-kpz}.

Other issues  worth remarking are that the behaviors of \(\langle\Phi[h]\rangle\) and \(\langle h\rangle\) are correlated, and the NEP's time behavior---both with and without external forcing---indicates a (non Kramers) activation-like phenomenon. In addition, all the analysis made above for 1d can be easily extended to higher dimensions.

We have discussed some strengths of a novel tool for describing the KPZ dynamics, the NEP approach. This innovative framework is expected to contribute answering questions that remain open by today, and handling the KPZ problem from still another perspective.
\vskip6pt

Support by MINECO (Spain), under project No.\ FIS2014-59462-P, is acknowledged by HSW, MAR and RG; support by CONICET, UNMdP and UNC (Argentina), by RRD and JAR. The authors thank J.M. L\'opez, R. Toral, C. Escudero, P. Colet and E. Hern\'andez-Garc\'{\i}a for fruitful discussions.

\end{document}